\documentclass[prb,twocolumn,superscriptaddress,showpacs,floatfix]{revtex4}

\usepackage{amsfonts}

\usepackage{epsfig,amsmath,amssymb,latexsym}

\usepackage{subfigure}

\begin{document}

\title{Competing topological and Kondo insulator phases on a honeycomb lattice}
\preprint{1}

\author{Xiao-Yong Feng}
  \affiliation{Condensed Matter Group,
  Department of Physics, Hangzhou Normal University, Hangzhou 310036, China}
\affiliation{Chen Jian Gong Institute for Advanced Study, Hangzhou
Normal University, Hangzhou 310036, China}
\author{Jianhui Dai}
  \affiliation{Condensed Matter Group,
  Department of Physics, Hangzhou Normal University, Hangzhou 310036, China}
\affiliation{Chen Jian Gong Institute for Advanced Study, Hangzhou
Normal University, Hangzhou 310036, China}
\affiliation{Department
of Physics, Zhejiang University, Hangzhou 310027, China}

\author{Chung-Hou Chung}
  \affiliation{Electrophysics Department, National Chiao-Tung
  University, HsinChu, Taiwan, 300 R.O.C.}
\affiliation{National Center for Theoretical Sciences, HsinChu,
Taiwan, 300 R.O.C.}

\author{Qimiao Si}
  \affiliation{Department of Physics and Astronomy, Rice University,
Houston, TX 77005, USA }


\begin{abstract}
We investigate the competition between the spin-orbit interaction of
itinerant electrons and their Kondo coupling with local moments
densely distributed on the honeycomb lattice. We find
that the model at half-filling
displays a quantum phase transition between topological and Kondo
insulators at a nonzero Kondo coupling. In the Kondo-screened case,
tuning the electron concentration can lead to a new topological
insulator phase. The results suggest that  the heavy-fermion phase
diagram contains a new regime with a competition among topological,
Kondo-coherent and magnetic states, and that the regime may be
especially relevant to Kondo lattice systems with $5d$-conduction
electrons. Finally, we discuss the implications of our results in
the context of the recent experiments on SmB$_6$ implicating
the surface states of a topological insulator, as well as the existing experiments
on the phase transitions in SmB$_6$ under pressure and in
CeNiSn under chemical pressure.
\end{abstract}

\pacs{71.10.-w,71.27.+a,73.43.Nq,75.70.Tj} \maketitle

Systems containing both itinerant electrons and local moments
continue to attract intensive interests in modern condensed matter
physics. The antiferromagnetic exchange coupling between the two
components gives rise to the Kondo singlet ground state.
Historically, the Kondo effect in a single local-moment impurity
provided the understanding of the resistivity minimum in metals as
well as the physics of dilute magnetic alloys and quantum
nanostructures \cite{Hewson, Coleman1}. In the concentrated case,
consideration of the Kondo effect and its competition with
magnetically ordered ground states has been playing a central role
in the understanding of the novel phases and quantum criticality of
heavy fermion materials \cite{Si1}. For the half-filled limit of the
Kondo lattice system, the Kondo effect gives rise to the
paramagnetic Kondo insulator (KI)
state\cite{Coleman1,Aeppli-Fisk,Ueda, Riseborough}.

Recently, the quantum spin Hall insulator in two dimensions (2D) and
the topological insulator (TI) more generally have attracted
extensive interest\cite{Hasan-Kane,Qi-Zhang}. These insulators have
a charge excitation gap in the bulk, but support gapless surface
states protected by time-reversal symmetry (TRS). The surface states
constitute a helical liquid where the spin orientation is locked
with the direction of electron momentum\cite{Kane-Mele1,WBZ}.
Although they are robust against weak disorders that preserve TRS,
the surface states may be influenced by magnetic impurities. For
example, the conductance of 1D edge helical liquid of a 2D TI in the
presence of a single magnetic impurity can exhibit a logarithmic
behavior at high temperatures and goes to the unitarity limit at
$T=0$ due to the formation of a Kondo singlet \cite{WBZ,
Maciejko-Zhang}. This is in contrast to the Kondo problem in
conventional Luttinger liquids, where even very weak Coulomb
interaction leads to vanishing conductance at zero
temperature\cite{Kane-Fisher}.
Generally speaking, the Kondo screening of magnetic impurities on
the surface of TI's may not necessarily be complete due to the
$SU(2)$ breaking of the spin-orbit coupling(SOC)\cite{Feng-Zhang},
and the effective RKKY interaction between the local moments can be
mediated by the edge carries, leading to an in-plane noncollinear
and helical
order\cite{Gao-Zhang,Biswas-Balatsky,Garate-Franz,Zhu-Chang}.
For magnetic impurities in TI's, previous studies have focused on
the effect of {\it surface impurities}, i.e., magnetic impurities
positioned on the surface of TI's,  or coupled effectively to the
surface states\cite{Maciejko}. Whether and how the {\it bulk
magnetic impurities} influence the properties of TI's is largely an
open problem.

From the perspective of heavy-fermion physics, very interesting
properties are emerging from materials which involve $5d$ electrons,
such as the pyrochlore Pr$_2$Ir$_2$O$_7$ \cite{nakatsuji}. The
significant SOC of the $5d$ electrons may give rise to topologically
non-trivial physics for the $5d$ electrons alone, raising the
intriguing question of the interplay between topological and Kondo
physics. The regime of transitions among the competing ground states
represents a setting in which the effects of strong interactions on
TI's may become more tractable. Furthermore, Kondo insulators
themselves may become topological as a result of the symmetry
properties of the hybridization matrix \cite{Dzero-Coleman}.

\begin{figure}[t!]
\includegraphics [width=7cm]{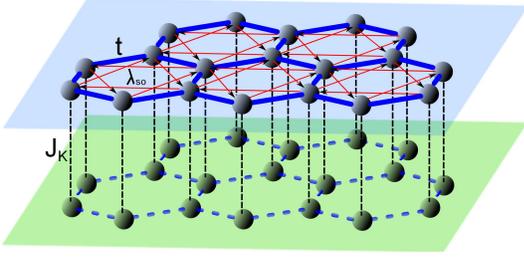}
\caption{Itinerant electrons moving on a honeycomb lattice while
coupled vertically to the local spins on a parallel lattice.}
\label{system}
\end{figure}

Motivated by these recent developments, in this letter we study a
dense set of magnetic local moments
interacting with the spin-orbit
coupled itinerant electrons on the honeycomb lattice as illustrated in  Fig.
\ref{system}. Such a system is relevant for the graphene/magnetic
moment interface and could be constructed through cold atoms in an
optical lattice. The system could also be realized by growing a 2D
TI on an appropriate magnetic insulating substrate; the similar
heterostructures involving TI Bi$_2$Se$_3$ thin films and
superconducting layers have already been fabricated by the molecular
beam epitaxy technique\cite{Jia}. It may very well be built based on
the existing  $5d$ electron based iridates on the honeycomb lattice,
such as Na$_2$IrO$_3$ \cite{gegenwart}.
Finally, given that recent experiments in SmB$_6$ have provided tentative evidence
 for the surface states of a topological
insulator
\cite{Wolgast,Botimer}, our results here on the transitions
between topological insulator and Kondo coherent states lead to the intriguing
question of what happens to such surface states when
SmB$_6$ and related intermetallic systems are tuned by
external or chemical pressure (see below).

The model we consider, illustrated in Fig.\ref{system}, is specified by the Hamiltonian
\begin{eqnarray}\label{Kondo}
H&=&-t\sum_{\langle\textbf{i}\textbf{j}\rangle\sigma}c^{\dagger}_{\textbf{i}\sigma}
c_{\textbf{j}\sigma} + i\lambda_{so}\sum_{\ll
\textbf{i}\textbf{j}\gg\sigma\sigma'}v_{\textbf{i}\textbf{j}}c^{\dagger}_{\textbf{i}\sigma}
{\sigma}^z_{\sigma\sigma'}
c_{\textbf{j}\sigma'}\nonumber\\
&&+J_K\sum_{\textbf{i}} {\vec s}_{\textbf{i}}\cdot {\vec
S}_{\textbf{i}},
\end{eqnarray}
where $c_{\textbf{i}\sigma}$ annihilates an electron at site
$\textbf{i}$ with spin component $\sigma=\uparrow, \downarrow$,
${\vec s}_{\textbf{i}}=c^{\dagger}_{\textbf{i}\sigma}({\vec
\sigma}_{\sigma\sigma'}/2)c_{\textbf{i}\sigma'}$,  and ${\vec
S}_{\textbf{i}}$ represents the local moments with ${\vec \sigma}$
being the Pauli matrices. The parameters $t$ and $\lambda_{so}$ are
the nearest neighbor hopping energy and the next-nearest-neighbor
{\it intrinsic} (Dresselhaus-type) SOC of the conduction electrons
respectively, with $v_{\textbf{i}\textbf{j}}=\pm1$ depending on the
direction of hopping between the next-nearest-neighbor sites.
Finally, $J_K$ is the {\it antiferromagnetic} Kondo coupling between
the spins of conduction electrons and local impurities. The model
Eq.(\ref{Kondo}) minimally interpolates the Kane-Mele Hamiltonian
($J_K=0$)\cite{Kane-Mele1,Kane-Mele2} and the standard Kondo lattice
Hamiltonian ($\lambda_{so}=0$).
We note that recent studies have focused on the effect of Hubbard
$U$ interaction of the conduction electrons \cite{Rachel-Hur,
HLA,YXL,Lee,ZZW,Quan}.

To proceed, we note that the model of Eq.(\ref{Kondo}) is connected
to the Anderson lattice Hamiltonian
\begin{eqnarray}\label{Anderson}
H=H_{KM}+H_{cd}+H_d,
\end{eqnarray}
where $H_{KM}$ is the Kane-Mele Hamiltonian [the first two terms of
Eq.(1)], $H_{cd}=V\sum_{\textbf{i}\sigma}
(c^{\dag}_{\textbf{i}\sigma}d_{\textbf{i}\sigma}+h.c.)$ is the
hybridization between the itinerant electrons and localized
$d$-electrons, and
$H_d=E_0\sum_{\textbf{i}\sigma}d^{\dag}_{\textbf{i}\sigma}d_{\textbf{i}\sigma}+
U\sum_{\textbf{i}}n_{d\textbf{i}\uparrow}n_{d\textbf{i}\downarrow}$
is for the local electrons with $E_0$ being the local energy level
and $U$ the on-site Coulomb repulsion of local electrons. The models
described by Eqs.(\ref{Kondo}) and (\ref{Anderson}) are equivalent
provided that, in the absence of SOC, the $d$-electrons are in the
Kondo regime ($U$ is sufficient large and $E_0$ is well below the
Fermi energy ($E_F$) of the conduction band). In this regime,
$J_K\sim V^2[\frac{1}{E_F-E_0}+\frac{1}{U-E_F+E_0}]$.
Our calculations will be carried out in Eq.(\ref{Anderson}). As our
focus is on the competition between the TI and KI at half filling,
we shall mainly consider the paramagnetic states.
\begin{figure}[t!]
\includegraphics [width=6cm]{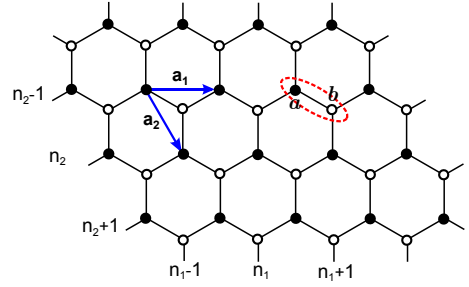}
\caption{The unit cell and primitive translation vectors.}
\label{lattice}
\end{figure}

In the momentum ${\bf k}$-space, the conduction electron part of
Hamiltonian takes the form of
$H_{KM}=\sum_{\textbf{k}\sigma}C_{\textbf{k}\sigma}^{\dag}M_{\textbf{k}\sigma}C_{\textbf{k}\sigma}$,
 with $C_{\textbf{k}\sigma}^{\dag} =
(c_{a,\textbf{k}\sigma}^{\dag},c_{b,\textbf{k}\sigma}^{\dag})$ and
\begin{eqnarray}
M_{\textbf{k}\sigma}=\left(\begin{array}{cc}
                                             \sigma \Lambda _{\textbf{k}}-\mu & \epsilon_{\textbf{k}} \\
                                           \epsilon_{\textbf{k}}^* &  -\sigma \Lambda_{\textbf{k}}-\mu \\
                                         \end{array}
                      \right),
\end{eqnarray}
where,  $\sigma=+1$ and $-1$ refers to spin up and spin down,
$\Lambda_{\textbf{k}}=2\lambda_{so}[\sin{k_1}-\sin{k_2}-\sin{(k_1-k_2)}]$,
$\epsilon_{\textbf{k}}=-t(1+e^{-ik_1}+e^{-ik_2})$. We have included
the chemical potential $\mu$-term to control the electron filling.
The subscripts $a$ and $b$ denote two sublattices of the honeycomb
lattice as shown in Fig.\ref{lattice}. Each unit cell has two
adjacent a, b sites, and the primitive vectors are $\textbf{a}_1$
and $\textbf{a}_2$.

For the local electrons, we consider the large-$U$ limit and utilize
the slave-boson method \cite{Hewson}.
The local electrons are expressed as
$d^{\dag}_{\textbf{i}\sigma}=f^{\dag}_{\textbf{i}\sigma}b_{\textbf{i}}$,
with $f^{\dag}_{\textbf{i}\sigma}$ and $b_{\textbf{i}\sigma}$ being
respectively fermionic and bosonic operators satisfying the
constraint $b^{\dag}_{\textbf{i}}b_{\textbf{i}}+\sum_{\sigma}
f^{\dag}_{\textbf{i}\sigma}f_{\textbf{i}\sigma}=1$. Introducing the
basis
$\Psi^{\dagger}_{\textbf{k}\sigma}=(c_{a,\textbf{k}\sigma}^{\dag},
c_{b,\textbf{k}\sigma}^{\dag},f_{a,\textbf{k}\sigma}^{\dag},f_{b,\textbf{k}\sigma}^{\dag})$
in the $\bf k$-space, the  mean-field Hamiltonian is expressed as
$H_{MF}
=\sum_{\textbf{k}\sigma}\Psi_{\textbf{k}\sigma}^{\dag}H_{\textbf{k}\sigma}\Psi_{\textbf{k}\sigma}$
with
\begin{eqnarray}\label{Matrix}
H_{\textbf{k}\sigma}=\left(
                       \begin{array}{cc}
                         M_{\textbf{k}\sigma} & rV\cdot I \\
                         rV\cdot I & (E_0+\lambda)\cdot I \\
                       \end{array}
                     \right).
\end{eqnarray}
Here, $I$ is a $2\times 2$ identity matrix, $r=\langle b\rangle$ is
the condensation density of the bosons, and $\lambda$ is the
Lagrange multiplier introduced to implement the constraint. We will
carry out our calculations for $N=2$ ($\sigma=\pm 1$); a large-$N$
generalization in the presence of SOC may also be considered
\cite{Dzero}. The quasiparticle bands of the mean-field Hamiltonian
Eq. (\ref{Matrix}) are degenerate for the two spin components. For
each spin component, the Hamiltonian can be diagonalized (even
though the matrix is $4 \times 4$) giving rise to the quasiparticle
dispersion
\begin{eqnarray} \begin{array}{c}
                   E^{(1)}_{\textbf{k}} =
\frac{1}{2}\left(G_{\textbf{k}+}+\sqrt{G^{2}_{\textbf{k}-}+4r^2V^2}\right)-\mu \\
                   E^{(2)}_{\textbf{k}} =
\frac{1}{2}\left(G_{\textbf{k}-}+\sqrt{G^{2}_{\textbf{k}+}+4r^2V^2}\right)-\mu \\
                    E^{(3)}_{\textbf{k}} =
\frac{1}{2}\left(G_{\textbf{k}+}-\sqrt{G^{2}_{\textbf{k}-}+4r^2V^2}\right)-\mu \\
                   E^{(4)}_{\textbf{k}} =
\frac{1}{2}\left(G_{\textbf{k}-}-\sqrt{G^{2}_{\textbf{k}+}+4r^2V^2}\right)-\mu
                 \end{array}\label{energy}
\end{eqnarray}
with
$G_{\textbf{k}\pm}=E_0+\lambda+\mu\pm\sqrt{\Lambda^2_{\textbf{k}}+|\epsilon_{\textbf{k}}|^2}$.
The parameters $r$ and $\lambda$ are determined by the following equations
\begin{eqnarray}
\frac{1}{2{\cal N}}\sum_{\textbf{k}\sigma;\alpha=a,b}\langle
f^{\dag}_{\alpha,\textbf{k}\sigma}f_{\alpha,\textbf{k}\sigma}\rangle+ r^2 &=&1 \\
\frac{V}{2{\cal N}}\sum_{\textbf{k}\sigma;\alpha=a,b}\Re{\langle
c^{\dag}_{\alpha,\textbf{k}\sigma}f_{\alpha,\textbf{k}\sigma}\rangle}+r\lambda&=&0
\end{eqnarray}
with ${\cal N}$ being the total number of unit cells and $\Re$
indicating the real part. In the following we shall mainly
consider
the half-filled case, corresponding to $\mu=0$.

The formation of the quasiparticle bands, specified by
Eq.(\ref{energy}), requires the renormalized hybridization
$V^*=rV\neq 0$. By contrast, if ${V}^*=0$, the spectra separate into
the decoupled conduction bands and local level. Moreover, the band
inversion takes place at ${ V}^*=0$. While this feature is hidden
and not important in ordinary Kondo lattice problems, it is crucial
in the present problem because now the conduction bands are from the
TI. As a consequence, the bulk gap of TI closes at the onset of
$V^*$, leading to a quantum phase transition to the KI.
\begin{figure}[t!]
\includegraphics [width=9cm]{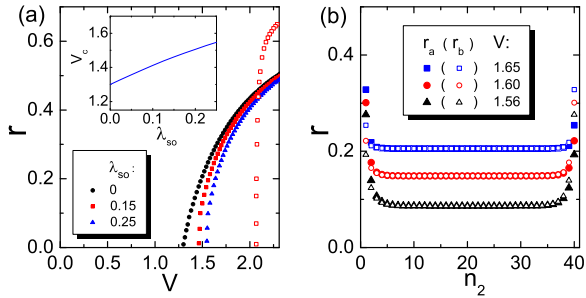}
\caption{(a)Large system with periodic boundary condition: The
mean-field parameter r as a function of V for $\lambda_{so}=0,
0.15$, and $0.25$. As a comparison, the red open square is for the
corresponding single-ion Kondo problem for $\lambda_{so}=0.15$. The
inset shows the critical $V_c$ as a function of $\lambda_{so}$. All
the coupling constants are in unit of $t$. (b) The site-dependence
of $r$ for the lattice with zig-zag edges for several $V$ in the
Kondo phase, the width $N_2=40$, $\lambda_{so}=0.15$.}
\label{parameter}
\end{figure}

At zero temperature,  $r$ ( or $V^*$ ) is non-zero only if $V$ is
larger than a critical $V_c$, as a result of the suppressed density
of conduction electron states for a honeycomb lattice.
Fig.\ref{parameter}(a) shows the numerical results for the
$V$-dependence of $r$ for several values of $\lambda_{so}$. The
local level $E_0$ is taken at the bottom of the conduction band. The
critical $V_c\sim 1.3$ for $\lambda_{so}=0$, and increases almost
linearly with $\lambda_{so}$, as seen in the inset of
Fig.\ref{parameter}(a). When $V<V_c$, $r=0$, indicating the Kondo
destruction, so the system remains in the TI phase with a bulk gap
$\Delta_{T}=6\sqrt{3}\lambda_{so}$. While for $V>V_c$, $r\neq0$, the
Kondo screening emerges and the band inversion takes place
immediately, so the system enters into the KI phase. For small $r$,
the KI phase has a finite hybridization gap $\Delta_{K}\sim
2r^2V^2/3t$. This is the direct band gap at the $\Gamma$-point where
the contribution from the SOC vanishes.

It is interesting to compare the results here for the Kondo-lattice
problem with those for its counterpart of a single ion magnetic
impurity imbedded in the bulk of the 2D TI. Using the same method,
and for $\lambda_{so}=0.15$ as an example shown in
Fig.\ref{parameter}(a), we find $V_c\sim 2.07$ for the single ion
Kondo screening which is much larger than $V_c\sim 1.45$ determined
here. In the absence of SOC, the finite $V_c$ is due to the fact
that the electron host is a pseudo gap system so that the single ion
Kondo screening needs a nonzero Kondo coupling comparable to the gap
amplitude\cite{Withoff,Ingersent,Neto}. The enhancement of Kondo
effect comparing to the single ion Kondo screening is similar to the
Kondo lattice with $d$-wave superconducting conduction
electrons\cite{Sheehy-Schmalian}.



We next investigate the surface states of the finite system with
boundaries. We take a 2D ribbon by cutting two zizag edges with
width $N_2$, while the size along $\textbf{a}_1$ remains infinite.
Then the boson mean-field $r$ is dependent on the coordinate $n_2$
and the sublattices, and can be denoted by $r_a(n_2)$ and $r_b(n_2)$
respectively. We have $r_b(n_2)=r_a(N_2-n_2)$ due to the inversion
symmetry. Fig.\ref{parameter}(b) shows the site-dependence of $r_a$
and $r_b$ for $N_2=40$ and $\lambda_{so}=0.15$. A general feature is
that $r$ decreases rapidly from the edge to the bulk. This feature
is attributed to the gapless edge states. $r(n_2)$ is almost flat
away from the edges ($5<n_2<35$) as shown in Fig.\ref{parameter}(b),
indicating that the finite size effect is relatively small for
$N_2=40$.
\begin{figure}[t!]
\includegraphics [width=8.5cm]{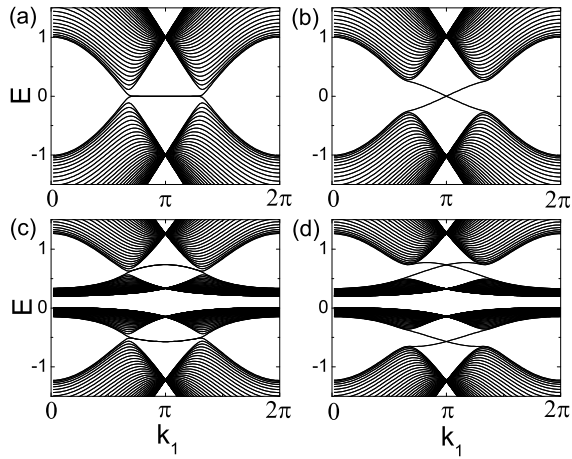}
\caption{The energy spectra for a ribbon of width $N_2=40$. (a)$V=0$ and
$\lambda_{so}=0$; (b)$V=0$ and $\lambda_{so}=0.03$; (c)$V=1.7$ and
$\lambda_{so}=0$; (d)$V=1.7$ and $\lambda_{so}=0.03$;}
\label{spectra}
\end{figure}

Figure \ref{spectra} shows the energy spectra with four sets of
parameters; here, $\mu=0$ is imposed. Figs. \ref{spectra}(a) and (b)
display the spectra of the conduction electrons in the absence of
Kondo singlet (${V}^*=0$) and for $\lambda_{so}=0,0.03$,
respectively. The edge states with a single Dirac point at the Fermi
energy in Fig.\ref{spectra}(b) manifest the TI phase
\cite{Kane-Mele1,Kane-Mele2}. In comparison, Figs. \ref{spectra} (c)
and (d) are the spectra for $V>V_c$. The Kondo-singlet formation is
clearly reflected in the hybridization gap at half-filling and the
relatively narrow flat bands near the Fermi energy (near the
transition point the flatness is measured by $t/V^*$ ).

Furthermore, we observe that in the KI phase, the narrow bands can be
separated from the continuum by increasing the SOC, leading to a
bulk gap at the 1/4- or 3/4- fillings (achieved by tuning the
chemical potential $\mu\neq 0$). This
is the direct band gap between $E^{(3)}_{{\bf k}}$ and
$E^{(4)}_{{\bf k}}$ or between $E^{(1)}_{{\bf k}}$ and
$E^{(2)}_{{\bf k}}$ at the points $(2\pi/3,4\pi/3)$ and
$(4\pi/3,2\pi/3)$, respectively, with the gap magnitude
$\Delta_{KT}\sim 6{\sqrt 3}r^2V^2\lambda_{so}/\mu^2$ for the large
bulk system. Moreover, for the finite system with boundaries, the
edge states emerge again with a Dirac point within the bulk gap.
This feature is robust
for a range of the chemical potential
$\mu$
corresponding to the 1/4- or 3/4- filling.
We have also calculated
the $\mathbb{Z}_2$ topological bulk invariant $\Xi_{2D}$ following
the monodromy approach developed in Ref. \cite{Prodan}. The result
confirms that $\Xi_{2D}=1$ at half-filling and $\Xi_{2D}=-1$ at 1/4-
or 3/4-filling. Therefore, the insulating phase at 1/4- or
3/4-filling is topologically non-trivial, and its surface states,
while having the spin direction locked by the momentum, contain the
contributions from both conduction and local electrons.
Hence in the case of Fig.\ref{spectra}(d) we have a new TI phase
with the Kondo-singlet formation and a surface heavy-fermion helical
liquid.

We now consider the TI-KI transition around $V_c$. In the present
analysis at the saddle-point level,
the onset of Kondo effect
is continuous (Fig. \ref{parameter}(a)).
Correspondingly,
 the KI
gap sets in continuously. By contrast, on the TI side there is
simply a decoupling of the conduction-electron and local-moment
components.
However,
we show that the quantum fluctuations beyond the saddle point
reduce the bulk
gap from the TI side,
as is described in some detail in the Supplementary Materials (SM).
The situation is similar to the case of single-impurity
pseudo-gapped Kondo problem,
for which numerical renormalization group
calculations, for instance, establish a well-defined second-order
phase transition for the Kondo-destruction quantum critical point
\cite{ingersent-si-prl2002}. In the lattice case, the RKKY
interaction between the local moments, which is mediated by Kondo
coupling in our model, will also induce magnetic order.
Taking into account the magnetic order will leave the KI phase intact;
as is standard, the Kondo screening present in the KI phase quenches
the local moments and their ordering tendency.
In the TI phase,
we have explicitly shown (in the SM) that an antiferromagnetic order, characterized
by the order parameter $M$,
will reduce the TI bulk gap to $\Delta_T=2(3\sqrt 3 \lambda_{so}-J_K
M)(1-V^2/E_0^2)$.
This TI gap remains non-zero for a finite range of $M$ and $V$; in other words,
the TI phase remains stable in the presence of an antiferromagnetic order for
a range of $V<V_c$.
Our results can be understood based on general arguments:
in the presence of magnetic order,
the $Z_2$ topological
invariant is replaced
by two spin-Chern numbers which remain
unchanged when the time reversal symmetry is broken by the magnetic
order\cite{LSheng06,Prodan09}. Meanwhile, the surface states remain
gapless unless the bulk gap closes\cite{LSheng11,LSheng12}.

While a detailed transitions among  KI and TI phases on the one hand,
and antiferromagnetic
order on the other is beyond the scope of the present work,
our work does reveal that
the heavy-fermion phase
diagram contains a hitherto unexplored new regime
with a competition among topological,
Kondo-coherent and magnetic states; such competition involves
the physics of Kondo destruction and associated
local quantum criticality \cite{lqcp}.
%
In other words,
when the magnetic order
and related dynamical effects are
incorporated in our analysis, the TI-KI transition discussed here
will represent a regime where topological effects strongly interplay
with the onset of magnetism and Kondo coherence. The simplification
that proximity to quantum criticality brings may very well make the
interaction effects on the TI phase and its associated surface
states  more tractable.

We close by noting that we have treated the
hybridization to be ${\bf k}$-independent. When the spin- and ${\bf
k}$-dependences of the hybridization is incorporated, part of the KI
phase may itself become topological, as emphasized by Ref.
\cite{Dzero-Coleman}.


We now briefly discuss our work in the context of 4$f$-electron-based
Kondo insulators.
In SmB$_6$, it is known that a sufficiently large external pressure collapses
the Kondo coherence in SmB$_6$ and turns it into an antiferromagnetic
metallic state
 \cite{Bar05.1}.
Combined with the recent experimental evidence in
SmB$_6$ for
the edge states of a topological
insulator
\cite{Wolgast,Botimer}, this is
reminiscent of the transition among the topological and Kondo coherent/magnetic
states implicated by the present study. An intriguing question then arises, which deserves
the study of future experiments: what happens
to the candidate chiral edge states when SmB$_6$ is placed under  external pressure?
Along a similar line, CeNiSn is another intermetallic system believed to be a Kondo
insulator.
In CeNiSn, (negative) chemical pressure achieved through Pd- or Pt- substitution
for Ni is known to induce a transition out of its Kondo insulator state \cite{Sak92.1,Adr96.1,Kal00.1,Kas91.1}. It will therefore be informative to explore surface
states in
the
Ce(Pt$_{1-x}$Ni$_x$)Sn and
 Ce(Pd$_{1-x}$Ni$_x$)Sn series. Finally, it is worth noting that CePtSn has the distinction
in that it involves 5$d$ electrons with a large SOC.

To summarize, we have considered the effect of SOC of the conduction
electrons in a Kondo-lattice system. Our study offers the first
qualitative understanding of the competition between topological and
Kondo insulator ground states on a simple and yet generic model in
two dimensions.
While our analysis has so far been primarily confined to
the paramagnetic cases, our results already suggest that the overall
phase diagram of heavy-fermion systems includes a new regime with
competition among topological, Kondo-coherent and magnetic states.
This regime should be particularly prominent for heavy fermion
systems whose conduction electron band is associated with the
strongly spin-orbit-coupled $5d$ electrons. As such, our work opens
up a new regime of physical interest for compounds based on iridium, platinum
and related $5d$ elements.

We would like to thank C. Cao, P. Goswami, E. Morosan, A.
Nevidomskyy, and Y. Zhou for useful discussions. In particular, we
thank S. Paschen for discussions on the Kondo-insulator materials.
 This work was supported in
part by the NSF of China, the NSF of Zhejiang Province, the 973
Project of the MOST, the NSF Grant No.DMR-1006985, and the Robert A.
Welch Foundation Grant No.C-1411. C.H.C. acknowledges support by the
NSC Grant No.98-2918-I-009-06, No.98-2112-M-009-010-MY3, the
NCTU-CTS, the NCTS, the MOE-ATU program of Taiwan.

\widetext
\newpage

\section*{Supplementary Material --
Competing topological and Kondo insulator phases on a honeycomb
lattice}

by: Xiao-Yong Feng,  Jianhui Dai, Chung-Hou Chung, Qimiao Si

\vskip 1.0 cm

In this Supplementary Material, we first derive an effective action
of the Kondo lattice model with SOC by taking into account the
quantum fluctuations of the slave bosons. We show that while the
quantum fluctuations do not affect the Kondo insulator (KI) phase,
they do reduce the bulk gap of the topological insulator (TI) phase.
We then consider the effect of the magnetic order of the local
spins, which breaks the time-reversal symmetry. We show that the
magnetic order does not change the nature of the TI phase. The
notations here follow those introduced in the main text.

\subsection{A. Effect of quantum fluctuations beyond the saddle point}
We start
from expressing the Hamiltonian introduced in the main text
in the slave-boson representation,
\begin{eqnarray}
H
=&-&t\sum_{\langle\textbf{i}\textbf{j}\rangle\alpha}c_{\textbf{i}\sigma}^{\dag}c_{\textbf{j}\sigma}
+
i\lambda_{so}\sum_{\ll\textbf{i}\textbf{j}\gg\sigma\sigma'}v_{\textbf{i}\textbf{j}}c^{\dag}_{\textbf{i}\sigma}s^{z}_{\sigma\sigma'}c_{\textbf{j}\sigma'}-\mu\sum_{\textbf{i}\alpha}c_{\textbf{i}\sigma}^{\dag}c_{\textbf{i}\sigma}\\
 &+& V\sum_{\textbf{i}\sigma}
(c^{\dag}_{\textbf{i}\sigma}b^{\dag}_{\textbf{i}}f_{\textbf{i}\sigma}+f^{\dag}_{\textbf{i}\sigma}b_{\textbf{i}}c_{\textbf{i}\sigma})+(E_0+\lambda)\sum_{\textbf{i}\sigma}f^{\dag}_{\textbf{i}\sigma}f_{\textbf{i}\sigma}+\lambda\sum_{\textbf{i}}
b^{\dag}_{\textbf{i}}b_{\textbf{i}}\\
=&&\sum_{\textbf{k}\sigma}(c^{\dag}_{a,\textbf{k}\sigma},c^{\dag}_{b,\textbf{k}\sigma})\left(
                                                                                                     \begin{array}{cc}
                                                                                                       -\mu+\sigma\Lambda_{\textbf{k}} & \epsilon_{\textbf{k}} \\
                                                                                                       \epsilon_{\textbf{k}}^{*} & -\mu-\sigma\Lambda_{\textbf{k}} \\
                                                                                                     \end{array}
                                                                                                   \right)\left(
                                                                                                             \begin{array}{c}
                                                                                                               c_{a,\textbf{k}\sigma} \\
                                                                                                              c_{b,\textbf{k}\sigma} \\
                                                                                                             \end{array}
                                                                                                           \right)\\
&+&\frac{V}{\sqrt{{\cal N}}}\sum_{\alpha\textbf{k}\textbf{q}\sigma}
(c^{\dag}_{\alpha,\textbf{k}-\textbf{q}\sigma}b^{\dag}_{\alpha,\textbf{q}}f_{\alpha,\textbf{k}\sigma}+f^{\dag}_{\alpha,\textbf{k}\sigma}b_{\alpha,\textbf{q}}c_{\alpha,\textbf{k}-\textbf{q}\sigma})+(E_0+\lambda)\sum_{\alpha\textbf{k}\sigma}
f^{\dag}_{\alpha,\textbf{k}\sigma}f_{\alpha,\textbf{k}\sigma}+\lambda\sum_{\alpha\textbf{q}}
b^{\dag}_{\alpha,\textbf{q}}b_{\alpha,\textbf{q}}
\end{eqnarray}
In the above expression, $\alpha = a$ or $b$ is the index for
sublattices, $\lambda$ is introduced to implement the no double
occupation of local electrons. In this formulism, the bosons can
accommodate zero modes, which
are denoted by
$b^{\dag}_{\alpha,\textbf{q}=0}=b_{\alpha,\textbf{q}=0}=r\sqrt{{\cal
N}}$. Hence a nonvanishing $r$ corresponds to the Bose-Einestein
condensation of the slave-bosons. Quantum fluctuations are
contributed mainly from the bosons with non-zero ${\bf q}$. The
effective action $S_{eff}$ for the system, defined by $Tr e^{-\beta
H}=\int {\cal D}c^{*}_{\alpha,{{\bf k}\sigma}}{\cal
D}c_{\alpha,{{\bf k}\sigma}} {\cal D}f^{*}_{\alpha,{{\bf
k}\sigma}}{\cal D}f_{\alpha,{{\bf k}\sigma}} {\cal
D}b^{*}_{\alpha,{{\bf q}}}{\cal D}b_{\alpha,{{\bf q}}} e^{-S_{eff}}
$, is then given by
\begin{eqnarray}
S_{eff} =&&  S_c+
\sum_{\alpha \textbf{k}\omega\sigma}f^*_{\alpha,\textbf{k}\sigma}(\omega)(-i\omega+E_0+\lambda)f_{\alpha,\textbf{k}\sigma}(\omega)+\sum_{\alpha \textbf{q}\neq0,\Omega}b^*_{\alpha,\textbf{q}}(-i\Omega+\lambda)b_{\alpha,\textbf{q}}\\
&&+ \left[V\sum_{\alpha
\textbf{k}\omega\sigma}\left(rc^{*}_{\alpha,\textbf{k}\sigma}(\omega)+\frac{1}{\sqrt{{\cal
N}\beta}}\sum_{\textbf{q}\neq0,\Omega}c^{*}_{\alpha,\textbf{k}-\textbf{q}\sigma}(\omega-\Omega)b^{*}_{\alpha,\textbf{q}}(\Omega)\right)f_{\alpha,\textbf{k}\sigma}(\omega)+c.c\right].
\end{eqnarray}
Here,
\begin{eqnarray}
S_c=\sum_{\textbf{k}\omega\sigma}(c^*_{a,\textbf{k}\sigma}(\omega),c^*_{b,\textbf{k}\sigma}(\omega))\left(
                                                                                                     \begin{array}{cc}
                                                                                                       -i\omega-\mu+\sigma\Lambda_{\textbf{k}} & \epsilon_{\textbf{k}} \\
                                                                                                       \epsilon_{\textbf{k}}^{*} & -i\omega-\mu-\sigma\Lambda_{\textbf{k}} \\
                                                                                                     \end{array}
                                                                                                   \right)\left(
                                                                                                            \begin{array}{c}
                                                                                                              c_{a,\textbf{k}\sigma}(\omega) \\
                                                                                                              c_{b,\textbf{k}\sigma}(\omega) \\
                                                                                                            \end{array}
                                                                                                          \right)
\end{eqnarray}
is the bare action of the itinerant electrons,
$\omega=\frac{(2n+1)\pi}{\beta}$ ( $\Omega=\frac{2n\pi}{\beta}$) are
the Matsubara frequencies for fermions (bosons). After integrating
out the fermionic (Grassman) fields
$f_{\alpha,\textbf{k}\sigma}(\omega)$ and
$f^*_{\alpha,\textbf{k}\sigma}(\omega)$, the effective action
becomes
\begin{eqnarray}
S =&& S_c+\sum_{\alpha
\textbf{q}\neq0,\Omega}b^*_{\alpha,\textbf{q}}(-i\Omega+\lambda)b_{\alpha,\textbf{q}}-\sum_{\alpha\textbf{k}\omega\sigma}\frac{V^2r^2}{-i\omega+E_0+\lambda}c^{*}_{\alpha,\textbf{k}\sigma}(\omega)c_{\alpha,\textbf{k}\sigma}(\omega)\\
&&-\frac{1}{{\cal
N}\beta}\sum_{\alpha\textbf{k}\omega\textbf{q}\neq0,\textbf{q}'\neq
0,\Omega,\Omega',\sigma}\frac{V^2}{-i\omega+E_0+\lambda}c^{*}_{\alpha,\textbf{k}-\textbf{q}\sigma}(\omega-\Omega)c_{\alpha,\textbf{k}-\textbf{q}'\sigma}(\omega-\Omega')b^{*}_{\alpha,\textbf{q}}(\Omega)b_{\alpha,\textbf{q}'}(\Omega').
\end{eqnarray}
We integrate out the bosonic fields and,
 to order of
$V^2$,
obtain the following effective action for the conduction electrons
\begin{eqnarray}
S_{eff} &=&S_c
-\sum_{\alpha\textbf{k}\omega\sigma}\frac{V^2r^2}{-i\omega+E_0+\lambda}c^{*}_{\alpha,\textbf{k}\sigma}(\omega)c_{\alpha,\textbf{k}\sigma}(\omega)\\
&&-\frac{1}{\beta}\sum_{\alpha\textbf{k}\omega\Omega\sigma}\frac{V^2}{(i\Omega
-\lambda)[i(\omega-\Omega)-E_0-\lambda]}c^{*}_{\alpha,\textbf{k}\sigma}(\omega)c_{\alpha,\textbf{k}\sigma}(\omega)+{\cal
O}(V^{4}).
\end{eqnarray}


Now using the
identities,
\begin{eqnarray}
\frac{1}{(i\Omega-\lambda)[i(\omega+\Omega)-E_0-\lambda]}&=&\frac{1}{i\omega-E_0}\left(\frac{1}{i\Omega-\lambda}
- \frac{1}{i(\omega+\Omega)-E_0-\lambda}\right),\\
\frac{1}{\beta}\sum_{\Omega}\frac{1}{i\Omega-\lambda}&=&-n_B(\lambda)=0,\\
\frac{1}{\beta}\sum_{\Omega}\frac{1}{i(\omega+\Omega)-E_0-\lambda}&=&n_{F}(E_0+\lambda),
\end{eqnarray}
we arrive at the following expression for the effective action
\begin{eqnarray}
S_{eff}
&=&\sum_{\textbf{k}\omega\sigma}(c^*_{a,\textbf{k}\sigma}(\omega),c^*_{b,\textbf{k}\sigma}(\omega))G_{\sigma}(\textbf{k},\omega)^{-1}\left(
                                                                                                            \begin{array}{c}
                                                                                                              c_{a,\textbf{k}\sigma}(\omega) \\
                                                                                                              c_{b,\textbf{k}\sigma}(\omega) \\
                                                                                                            \end{array}
                                                                                                          \right),
\end{eqnarray}
where the inverse of the Green's function is
\begin{eqnarray}
G_{\sigma}(\textbf{k},\omega)^{-1}=\left(
                                                                                                     \begin{array}{cc}
                                                                                                       -i\omega+\frac{n_F(E_0+\lambda)V^2}{i\omega-E_0}+\frac{V^2r^2}{i\omega-E_0-\lambda}-\mu+\sigma\Lambda_{\textbf{k}} & \epsilon_{\textbf{k}} \\
                                                                                                       \epsilon_{\textbf{k}}^{*} & -i\omega+\frac{n_F(E_0+\lambda)V^2}{i\omega-E_0}+\frac{V^2r^2}{i\omega-E_0-\lambda}-\mu-\sigma\Lambda_{\textbf{k}} \\
                                                                                                     \end{array}
                                                                                                   \right).
\end{eqnarray}

Since the bare energy level of the local $f$-electrons $E_0<0$ is
much lower than the Fermi energy, the poles of the Green's function
are essentially determined by the condition $E_0+\lambda\sim 0$ when
$r\neq 0$. In this case, one recovers all the mean-field results as
discussed in the main text, where the quasiparticle bands are
equivalently given by the poles of the Green's function. The Kondo
regime where $r\neq 0$ is determined by $V>V_c$, with $V_c$ being
self-consistently solved by the mean field equations. Quantum
fluctuations contributed from the term
$\frac{n_F(E_0+\lambda)V^2}{i\omega-E_0}$ are negligible. Therefore
we conclude that the existence of the KI phase and its properties
are robust to the quantum fluctuations of the slave bosons.

On the other hand, when $V<V_c$, or $r=0$, one has
$n_F(E_0+\lambda)=\frac{1}{2}$, then the poles of the Green's
function are given by
\begin{eqnarray} \omega_{1} &=&
\frac{1}{2}\left[-(\mu+\sqrt{\Lambda_{\textbf{k}}^2+|\epsilon_{\textbf{k}}|^2}-E_0)+\sqrt{(\mu+\sqrt{\Lambda_{\textbf{k}}^2+|\epsilon_{\textbf{k}}|^2}+E_0)^2+2V^2}\right]\\
\omega_{2} &=&
\frac{1}{2}\left[-(\mu-\sqrt{\Lambda_{\textbf{k}}^2+|\epsilon_{\textbf{k}}|^2}-E_0)+\sqrt{(\mu-\sqrt{\Lambda_{\textbf{k}}^2+|\epsilon_{\textbf{k}}|^2}+E_0)^2+2V^2}\right]\\
\omega_{3} &=&
\frac{1}{2}\left[-(\mu+\sqrt{\Lambda_{\textbf{k}}^2+|\epsilon_{\textbf{k}}|^2}-E_0)-\sqrt{(\mu+\sqrt{\Lambda_{\textbf{k}}^2+|\epsilon_{\textbf{k}}|^2}+E_0)^2+2V^2}\right]\\
\omega_{4} &=&
\frac{1}{2}\left[-(\mu-\sqrt{\Lambda_{\textbf{k}}^2+|\epsilon_{\textbf{k}}|^2}-E_0)-\sqrt{(\mu-\sqrt{\Lambda_{\textbf{k}}^2+|\epsilon_{\textbf{k}}|^2}+E_0)^2+2V^2}\right]
\end{eqnarray}

At the half filling, $\mu=0$, there are two quasiparticle bands near
the Fermi energy, $\omega_{1}$ and $\omega_{2}$. While the bands
$\omega_{3}$ and $\omega_{4}$ are well below the Fermi energy and
are always occupied. When $V=0$, one simply recovers the two
electron bands decoupled from the local spins, with a direct band
gap between $\omega_{1}$ and $\omega_{2}$ opens at the Dirac points,
$\frac{1}{2}\Delta_T=\sqrt{\Lambda_{\textbf{k}}^2+|\epsilon_{\textbf{k}}|^2}=3\sqrt{3}\lambda_{so}$.
When $V (<V_c)$ is small, the band structures do not change, while
the amplitude of the band gap is given by
$\Delta_T(1-\frac{V^2}{2E_0^2})$. Therefore, the hybridization $V
(<V_c)$ just reduces the bulk gap of the TI phase. Combined with the
result shown in Fig.3(a),
our results suggest that the bulk gap decreases continuously when
the quantum critical point is approached from either sides.
Therefore we argue that
quantum fluctuations turn the TI-KI transition
into a canonical continuous form, with vanishing gaps from both
sides.

\subsection{B. Effect of antiferromagnetic order of the local spins}

We turn consider an antiferromagnetic order of the local spins
induced by the RKKY interactions. Our main concern here is how such
an order, which breaks the time-reversal symmetry, influences the
topological order. To address this issue, it is adequate to consider
the effect of an effective staggered magnetic field associated with
the antiferromagnetic order. For definiteness, we consider the
antiferromagnetic order to correspond to the local spins staggered
on the sublattices a and b.
We use $M$ to represent the magnetization of the $f$-electrons per
site in the even (or odd) sublattice, which will induce a
conduction-electron polarization
by the Kondo interaction.
This will add an additional term The effective Hamiltonian is:
\begin{eqnarray}
H_{J}=\sum_{a{\bf k}\sigma}\sigma I_R M f^{\dagger}_{a,{\bf
k}\sigma}f_{a,{\bf k}\sigma}-\sum_{b{\bf k}\sigma}\sigma I_R M
f^{\dagger}_{b,{\bf k}\sigma}f_{b,{\bf k}\sigma}-\sum_{a{\bf
k}\sigma}\sigma J_K M c^{\dagger}_{a,{\bf k}\sigma}c_{a,{\bf
k}\sigma}+\sum_{b{\bf k}\sigma}\sigma J_K M c^{\dagger}_{b,{\bf
k}\sigma}c_{b,{\bf k}\sigma}
\end{eqnarray}
to the total Hamiltonian.
Here, $I_R$ is the RKKY interaction $\sim \rho_0J_K^2$ , which is
small for small hybridization.

We follow
the approach similar to what was described above, obtaining the
inverse Green's function in the TI phase ($V<V_c$)
\begin{eqnarray}
G_{\sigma}(\textbf{k},\omega)^{-1}=\left(
                                                                                                     \begin{array}{cc}
                                                                                                       -i\omega+\frac{n_F(E_0+\lambda+\sigma I_R M)V^2}{i\omega-E_0-\sigma I_R M}-\mu+\sigma(\Lambda_{\textbf{k}}-J_K M) & \epsilon_{\textbf{k}} \\
                                                                                                       \epsilon_{\textbf{k}}^{*} & -i\omega+\frac{n_F(E_0+\lambda-\sigma I_R M)V^2}{i\omega-E_0+\sigma I_R M}-\mu-\sigma(\Lambda_{\textbf{k}}-J_K M) \\
                                                                                                     \end{array}
                                                                                                   \right).
\end{eqnarray}
Because in the absence of $I_RM$, $(E_0+\lambda)$ is close to 0 at
half-filling, it is reasonable to replace $n_F(E_0+\lambda\pm\sigma
I_R
 M)$ by $n_F(\pm\sigma I_RM)$, and when $M$ is very small,
$\frac{n_F(E_0+\lambda \pm \sigma I_R M)V^2}{i\omega-E_0 \mp \sigma
I_R M}\sim \frac{n_F(\pm\sigma I_RM)V^2}{i\omega-E_0}$. Therefore,
the poles of the Green's function is given by
$$
[-\omega+\frac{n_F(\sigma
I_RM)V^2}{\omega-E_0}+\sigma(\Lambda_{k}-J_K M)]
[-\omega+\frac{n_F(-\sigma
I_RM)V^2}{\omega-E_0}-\sigma(\Lambda_{k}-J_K M)]=|\epsilon_{k}|^2.
$$

Obviously, the quasiparticle bands are similar to those of the
original TI phase. At the gap position, $\epsilon_k=0$,
$\omega-\frac{V^2 n_F(\pm \sigma I_RM )}{\omega-E_0}=\pm
\sigma(\Lambda_k-J_K M)$. The effect of the staggered fields ($M>0$)
is evident: while $M$ simply reduces the energy $\Lambda_k$, it also
effectively renormalizes $V^2\rightarrow  2n_F(-I_RM)V^2$. They both
reduce the direct bulk gap
at zero temperature to be $2(3\sqrt {3}\lambda_{so}-J_K
M)[1-\frac{V^2}{E^2_0}]$ (at the Dirac points). Because the magnetic
order breaks the time reversal symmetry (TRS), the spin degeneracy
of the surface states splits\cite{KaneMele105}. It is well-known
that the TI with a TRS-breaking magnetic field is characterized by
the spin Chern number\cite{Haldane88,XLQ06,Fukui07}, and the $Z_2$
characterization is recovered when $M\rightarrow 0$
\cite{LSheng06,Prodan09}. Hence the TI phase persists when $M\neq 0$
unless the bulk gap $\Delta$ closes at certain large and finite
hybridization (and, relatedly, $J_K M$). Therefore, we conclude that
for small
hybridization, the antiferromagnetic order
keeps the KI phase intact. This demonstrates the validity of the
main conclusions obtained from the SBMF method. For sufficient large
hybridization (and, relatedly, $J_K M$), however, the system enters
the antiferromagnetic
 phase
 in which the surface states may
be gapped\cite{LSheng11,LSheng12}. The band structures at that
regime are rich but beyond the scope of the present paper.

\end{document}